\documentclass[aps,twocolumn,notitlepage,superscriptaddress,nofootinbib,prl]{revtex4-1}
\usepackage{amsmath}
\usepackage{epsfig}
\usepackage{multirow}
\usepackage{slashed}
\usepackage{amsmath}

\def\bea#1\eea{\begin{align}#1\end{align}}
\newcommand{\nnu}{\nonumber\\}
\newcommand{\bef}{\begin{figure}[!htp]}
\newcommand{\eef}{\end{figure}}

\begin{document}
\title{A global extraction of the jet transport coefficient in cold nuclear matter }

\date{\today}

\author{Peng Ru}
%\email{p.ru@m.scnu.edu.cn}
\affiliation{Institute of Quantum Matter, South China Normal University, Guangzhou 510006, China}       

\author{Zhong-Bo Kang}
%\email{zkang@physics.ucla.edu}                   
\affiliation{Department of Physics and Astronomy, University of California, Los Angeles, California 90095, USA}
\affiliation{Mani L. Bhaumik Institute for Theoretical Physics, University of California, Los Angeles, California 90095, USA}

\author{Enke Wang}
\email{hxing@m.scnu.edu.cn}
\affiliation{Institute of Quantum Matter, South China Normal University, Guangzhou 510006, China}       

\author{Hongxi Xing}
\email{wangek@scnu.edu.cn}
\affiliation{Institute of Quantum Matter, South China Normal University, Guangzhou 510006, China}       

\author{Ben-Wei Zhang}
%\email{bwzhang@mail.ccnu.edu.cn}
\affiliation{Key Laboratory of Quark $\&$ Lepton Physics (MOE) and Institute of Particle Physics, Central China Normal University, Wuhan 430079, China}

\date{\today}         

\begin{abstract}
Within the framework of the generalized QCD factorization formalism, we perform the first global analysis of the jet transport coefficient ($\hat q$) for cold nuclear matter. The analysis takes into account the world data on transverse momentum broadening in semi-inclusive electron-nucleus deep inelastic scattering, Drell-Yan dilepton and heavy quarkonium production in proton-nucleus collisions, as well as the nuclear modification of the structure functions in deep inelastic scattering, comprising a total of 215 data points from 8 data sets. For the first time, we clarify quantitatively the universality and probing scale dependence of the nuclear medium property as encoded in $\hat q$. We expect that the determined parametrization of $\hat q$ in cold nuclear matter will have significant impact on precise identification of the transport property of hot dense medium created in heavy ion collisions.
\end{abstract}

\maketitle

%%%%%%%%%%%%%%%%%%%%%%%%%%%%%%%%%%%%%%%%%%%%%%%%
{\it \ Introduction. }\
The exploration of the inner structure and properties of nuclear medium is of fundamental importance in nuclear science. In relativistic heavy ion collisions, of particular interest is the jet transport coefficient $\hat q$. Physically, $\hat q$ represents the transverse momentum broadening per unit length of an incoming jet encountered in a hot and dense medium, and describes the interaction strength between the hard probe and nuclear medium. In recent years, jet transport coefficient $\hat q$ has become a standard quantity in searching for and to characterize the properties of quark-gluon plasma \cite{Burke:2013yra}. $\hat q$ is the key quantity in the study of jet quenching phenomena and parton energy loss, as it arises in all the theoretical descriptions of energetic probes in heavy ion collisions~\cite{Baier:1996sk,Chen:2011vt, Majumder:2011uk}, which is remarkable. 

So far there has been significant efforts and a lot of progress in extracting $\hat q$ for {\it hot dense medium} through measurements of jet quenching at Relativistic Heavy Ion Collider (RHIC) and the Large Hadron Collider (LHC) \cite{Burke:2013yra, Zhou:2019gqk, Xie:2019oxg, Ma:2018swx, Chen:2016vem, Andres:2016iys}. On the other hand, a comprehensive study of the transport property of {\it cold nuclear matter} is still lacking, which we set out to address in this letter. Significant theoretical progress also makes now an opportune time to perform a detailed study with the theoretical inputs. For example, for a long time, $\hat q$ has been assumed to be a constant in most studies~\cite{Luo:1993ui,Qiu:2003vd,Qiu:2004da}. However, this is inconsistent with recent theoretical developments, where renormalization group equations for $\hat q$ has been derived through explicit calculations of radiative corrections~\cite{Kang:2013raa,Kang:2014ela,Kang:2016ron,Blaizot:2014bha,Iancu:2014kga,Liou:2013qya}. In particular, complete next-to-leading order (NLO) calculations of transverse momentum broadening for real scattering processes have shown the universality and scale dependence of the twist-4 quark-gluon correlation functions for nucleus \cite{Kang:2013raa,Kang:2014ela,Kang:2016ron}, which in turn implies the universality and scale-dependence of the corresponding $\hat q$ in the nuclear medium. However, the obtained evolution equation therein is not closed, thus require computation at even higher orders, which is unfortunately formidable in current quantitative study of scale dependence of the medium property.  

In this letter, we aim at a data driven understanding of the medium property of cold nuclear matter based on well defined physical observables and rigorous QCD factorization formalism, and thus to address how/whether the experimental data provide useful information on the $\hat q$. It has long been recognized that the phenomenon of transverse momentum broadening, defined as the difference of averaged transverse momentum square between nuclear and hadronic collisions, can be treated as an excellent observable to probe the nuclear medium property and the QCD multiple scattering dynamics \cite{Qiu:2001hj,Qiu:2005ki}. Experimentally, significant transverse momentum broadening effects have been observed in various experiments involving different identified final state particles at different collision energies and collision systems \cite{Airapetian:2009jy,Alde:1991sw,Leitch:1995yc,Vasilev:1999fa,McGaughey:1999mq,Peng:1999gx,Adare:2012qf,Adam:2015jsa,Bordalo:1987cr}. Theoretically, this observable has been systematically calculated within the rigorous theoretical framework of generalized factorization formalism \cite{Qiu:1990xxa,Qiu:1990xy,Luo:1994np,Luo:1993ui,Luo:1992fz}, or high-twist expansion approach, in which one can attribute the transverse momentum broadening to the next-to-leading power correction in the momentum transfer, the size is determined by the twist-4 parton-parton correlation functions in nucleus. 

It has been shown that $\hat q$ in semi-inclusive electron-nucleus deep inelastic scattering (SIDIS) and Drell-Yan (DY) have the same functional form and satisfy the same QCD evolution equation. The experimental check of the universality of nuclear medium property would provide a critical test of the generalized factorization formalism. In this letter, we will carry out, for the first time, a combined fit of world data on transverse momentum broadening in SIDIS \cite{Airapetian:2009jy}, DY dilepton and heavy quarkonium production in proton-nucleus (pA) collisions \cite{Alde:1991sw,Leitch:1995yc,Vasilev:1999fa,McGaughey:1999mq,Peng:1999gx,Adare:2012qf,Adam:2015jsa}. Notice that the dynamical shadowing effect as observed in electron-nucleus (eA) deep inelastic scattering (DIS) is also sensitive to the value of $\hat q$ \cite{Qiu:2003vd}, therefore we include Fermilab E665 data \cite{Adams:1992nf,Adams:1995is} into our analysis as well. From our global analysis presented below, we provide the first quantitative evidence of the universality and scale dependence of the nuclear medium property. 

%%%%%%%%%%%%%%%%%%%%%%%%%%%%%%%%%%%%%%%%%%%
{\it \ Transverse momentum broadening and $\hat q$. }\
The nonperturbative but universal $\hat q$ can be accessed in measurements of transverse momentum broadening. Taking SIDIS as an example, the transverse momentum broadening is defined as $\Delta \langle p_T^2\rangle = \langle p_T^2\rangle _{eA} - \langle p_T^2\rangle_{ep}$, with $p_T$ the transverse momentum of final state hadron and $\langle p_T^2\rangle_{eA/ep}$ the average transverse momentum in $eA$ (or $ep$) collisions. The averaged transverse momentum square is an inclusive observable and perturbatively calculable as the transverse momentum is integrated over. The first nontrivial leading contribution comes from final state double scattering manifested as twist-4 power corrections to the cross section \cite{Guo:1998rd}
\bea
\Delta \langle p_T^2\rangle = \frac{4\pi^2\alpha_sz_h^2}{N_c}
\frac{\sum_q T_{qg}(x_B,0,0,\mu^2) D_{h/q}(z_h,\mu^2)}{\sum_q f_{q/A}(x_B,\mu^2) D_{h/q}(z_h,\mu^2)}\,,
\label{eq-dis}
\eea
where $f_{q/A}(x_B,\mu^2)$ is the parton distribution function with $x_B$ the Bjorken-$x$, and $D_{h/q}(z_h,\mu^2)$ is the hadron fragmentation function with $z_h$ the momentum fraction. The twist-4 quark-gluon correlation function is defined as
\bea
T_{qg}(x,0,0) =& \int\frac{dy^-}{2\pi}e^{ixp^+y^-} \int\frac{dy_1^-dy_2^-}{4\pi}\theta(y_2^-)\theta(y_1^--y^-) \nnu
&\times\langle A| \bar \Psi_q(0)\gamma^+ F^+_{\alpha}(y_2^-)F^{\alpha+}(y_1^-)\Psi_q(y^-)|A\rangle\,,
\eea
which contains the fundamental properties of the nuclear medium as probed by a propagating quark. Under the approximation of a large and loosely bound nucleus, one can neglect the momentum and spatial correlations of two nucleons. Thus $T_{qg}(x,0,0,\mu^2)$ can be effectively factorized as \cite{CasalderreySolana:2007sw}
\bea
T_{qg}(x,0,0,\mu^2) \approx \frac{9R_A}{8\pi^2\alpha_s} f_{q/A}(x,\mu^2) \hat q(x,\mu),
\label{eq:Tqg}
\eea   
where $R_A$ is the nuclear radius and $\hat q(x,\mu)$ is the nuclear geometry averaged quark jet transport coefficient. 

In Drell-Yan process, the transverse momentum broadening of final state dilepton is caused purely by initial state multiple scattering. The final expression analogous to Eq. (\ref{eq-dis}), except that one replaces the fragmentation functions with parton distribution functions for the beam proton \cite{Guo:1998rd,Fries:2002mu}. In heavy quarkonium production in pA collisions, the transverse momentum broadening receives contributions not only from the initial state multiple scattering analogy to that in Drell-Yan process, but also the final state double scattering between the heavy quark pair and nuclear medium. Details of the calculation and final expressions can be found in~\cite{Kang:2008us,Kang:2012am} for both the color evaporation model and non-relativistic QCD (NRQCD) effective theory. 

Last but not least, resummed power corrections to the DIS nuclear structure functions $F_2(x_B, Q^2)$ has been calculated systematically in the framework of the pQCD factorization approach with resummed high twist contributions \cite{Qiu:2003vd}. It is found that the shadowing effect as observed in experiment is also sensitive to the value of $\hat q$, therefore provide us with another type of good observable to probe the nuclear medium property. All these calculations are performed within the same collinear factorization framework, i.e., high-twist expansion.

%%%%%%%%%%%%%%%%%%%%%%%%%%%%%%%%%%%%%%%%%%%
{\it \ Global analysis and results. }
The idea of global analysis is to extract the nonperturbative functions entering the factorized cross sections. This technique has been extensively used to explore the nucleon 1D and 3D structures, in which parametrized forms of the nonperturbative functions are assumed for the global analysis. Similarly, we adopt the following flexible functional form to parametrize $\hat q$,
\bea
\hat q(x,\mu^2) = \hat q_0 \,\alpha_s(\mu^2) \,x^{\alpha}(1-x)^{\beta} \ln^{\gamma}(\mu^2/\mu_0^2)\,,
\label{eq-pramt}
\eea
where $\alpha_s(\mu^2)$ is introduced to offset the strong coupling constant $\alpha_s$ in the denominator of Eq.~\eqref{eq:Tqg}, and $\mu_0 = 1$~GeV is introduced to make the argument in the logarithm dimensionless. Thus, we have 4 free parameters $\hat q_0,~\alpha,~\beta$, and $\gamma$ to be fitted to experimental data. The term $\ln^{\gamma}(\mu^2/\mu_0^2)$ represents any deviation in the QCD evolution of $T_{qg}(x,0,0,\mu^2)$ from that of $f_{q/A}(x, \mu^2)$, see Eq.~\eqref{eq:Tqg}, and thus mimics the scale-dependence of $\hat q$ to be determined from the experimental data. In small-$x$ region, we expect $\hat q$ to be proportional to the gluon saturation scale $ Q_s^2 \propto x^{-1/3}$~\cite{GolecBiernat:1998js} and thus the factor $x^\alpha$ in~$\hat q$. Finally in the large-$x$ region, power corrections could also be different~\cite{Dokshitzer:1995qm,Brodsky:2000zu} and thus we have the factor $(1-x)^\beta$. We use the MINUIT package~\cite{James:1975dr} to perform a global fit of the $\hat q$ from world data. To be consistent with the region of applicability of collinear factorization formalism, we include only the data points with $Q^2 > 1$~GeV$^2$.

In this analysis, due to the lack of complete NLO calculations of transverse momentum broadening in eA and pA collisions, we stick to LO of pQCD results, where only diagonal twist-4 matrix elements are involved. We leave those involving off-diagonal twist-4 matrix element, such as energy loss calculations in eA \cite{Guo:2000nz,Zhang:2003yn} and pA collisions \cite{Xing:2011fb}, for future works when more data is available for reasonable constraints. As for the proton PDFs, we use CT14 at LO with $n_f=3$ active quark flavors \cite{Dulat:2015mca}. For Pion PDFs, we take the parametrization form as in Ref. \cite{Sutton:1991ay}. As for fragmentation functions, we use the DSS parametrization~\cite{deFlorian:2007aj}. As for heavy quarkonium, we set heavy quark mass $m_c=1.5~ {\rm GeV}, ~ m_b=4.5$ GeV. We set the renormalization and factorization scale the same $\mu_r^2=\mu_f^2=Q^2$, with $Q$ the invariant mass of the virtual photon or heavy quarkonium mass. The theoretical uncertainties due to leading-twist nonperturbative functions are largely canceled as the transverse momentum broadening is a ratio of transverse momentum weighted and the total cross sections. 

\begin{table}[th!]
\caption{\label{table1}
Data used in our analysis, the individual and total $\chi^2$ values of the fit. 
We employ cuts of $Q>1\,\mathrm{GeV}$ for the DIS structure function for reliable pQCD calculation.}
\begin{ruledtabular}
\begin{tabular}{lccc}
experiment & data type & data points & $\chi^2$ \\ \hline
HERMES \cite{Airapetian:2009jy}  & SIDIS     &    156  &  189.7  \\
FNAL-E772 \cite{Vasilev:1999fa}   & DY     &    4  &  1.6  \\
SPS-NA10 \cite{Bordalo:1987cr}    & DY     &   5  &  6.5  \\    
FNAL-E772 \cite{Alde:1991sw,Peng:1999gx} & $\Upsilon$     &   4  &  2.7  \\
FNAL-E866 \cite{Leitch:1995yc,McGaughey:1999mq} & $J/\psi$    &   4  &  2.4  \\
RHIC \cite{Adare:2012qf}   & $J/\psi$     &    10  &  31.0  \\
LHC  \cite{Adam:2015jsa}  & $J/\psi$    &    12  &  4.8  \\ 
FNAL-E665 \cite{Adams:1992nf,Adams:1995is}   & DIS  &  20  & 21.5
    \\ \hline
{\bf TOTAL:} & & 215 & 260.2  \\
\end{tabular}
\end{ruledtabular}
\end{table}
We now present results for our global analysis of SIDIS, DY, heavy quarkonium in pA and structure function in eA. In table I, we list all the data sets that are included in our analysis as well as their respective $\chi^2$ values with 4 free parameters in the fit. In total we have fitted 215 data points from 8 data sets. Shown in Fig. \ref{fig:xq} is the kinematic reach in existing data. The capability to probe $\hat q$ is mainly located in the intermediate $x_B$ and $Q^2$ region. Future measurements with wider kinematic coverage (e.g. at a future Electron Ion Collider) is indispensable for a complete understanding of medium property.
\bef
\psfig{file=qhat_distribute.eps, width=2.5in}
\caption{The range in Bjorken-$x$ and $Q^2$ accessible in existing data.}
\label{fig:xq}
\eef 

In order to clarify the $x$ and scale $\mu^2$ dependence of $\hat q(x, \mu^2)$, we follow the usual way to treat $\hat q$ as a constant in the fit, i.e., setting $\alpha=\beta=\gamma=0$ and eliminating $\alpha_s$ dependence in Eq.~\eqref{eq-pramt}. As expected and advocated above, with just one free parameter $\hat q_0$, the MINUIT fit fails to converge to reach a minimized $\chi^2$. To elaborate further, we perform the fit for individual data sets, and present the fitted $\hat q$ in each individual process in Fig.~\ref{fig:fit_const}. One can immediately conclude that the fitted $\hat q$ for different processes (or same process but in different kinematic regions) are not a single constant, and can even differ by a factor of two or more. Apparently this commonly used fit of constant $\hat q$ contradicts with the statement that $\hat q$ is universal as stated in Ref. \cite{Kang:2013raa}. This exercise thus hints that there would be nontrivial $x$ and $\mu^2$ dependence of $\hat q$. 
\bef
\psfig{file=qhat_fit_const.eps, width=3.0in}
\caption{The fitted $\hat q$ from each individual process with parameters $\alpha=\beta=\gamma=0$ in Eq. (\ref{eq-pramt}). ``F.", ``B." and ``C." represent for ``Forward", ``Backward" and ``Central" rapidity, respectively. Errors from the fits are indicated by the vertical bars.}
\label{fig:fit_const}
\eef

\begin{figure*}
\psfig{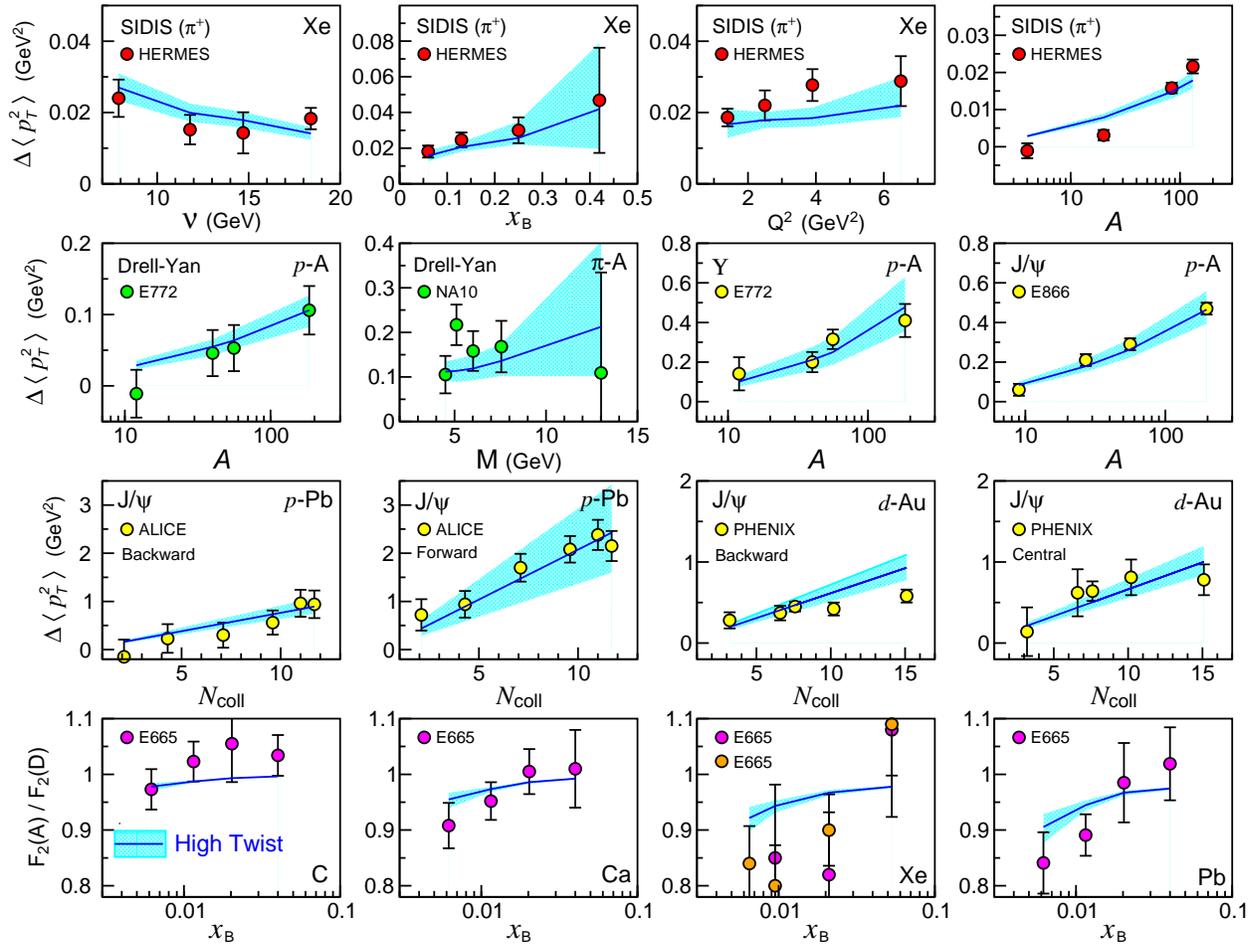}
\caption{Global fit compared to HERMES, FNAL, RHIC and LHC data, only some representative plots are shown.}
\label{fig:fit}
\end{figure*}

In this letter, we aim to overcome this unsatisfactory situation jeopardizing the reliability of generalized factorization formalism. We perform a global fit to all relevant high-quality data from eA and pA collisions with 4 free parameters. Specifically, the $x_B$, $\nu$, $Q^2$ dependence in SIDIS, DY, and heavy quarkonium would provide good constraints on the parametrization form of $\hat q$, and the atomic number $A$ distribution or number of binary collisions $N_{\rm coll}$ dependence will serve as a strong check of high twist expansion approach as it predicts that multiple scattering leads to linear medium size dependence of transverse momentum broadening. In Fig.~\ref{fig:fit}, we compare our theoretical results with some representative data sets, including transverse momentum broadening in SIDIS, transverse momentum broadening for DY lepton pair and heavy quarkonium in pA collisions, and nuclear modification ratio $F_2(A)/F_2(D)$ in the DIS structure function $F_2$. The solid curve corresponds to the best fit with $\hat q_0 = 0.02$ GeV$^2/$fm, $\alpha=-0.17,~ \beta=-2.79$, and $\gamma=0.25$, while the shaded area corresponds to the error in the global fit with $90\%$ confidence level. Standard Lagrange multiplier technique \cite{Stump:2001gu,Pumplin:2000vx} is employed in order to assess the uncertainty of the $\hat q$ determined in the fit. 

We observe from Fig. \ref{fig:fit} that the generalized factorization formalism can describe all data sets very well. This provides us with the first evidence about the universality of medium property as characterized by $\hat q$, and also identify quantitatively for the first time the kinematic $x$ and the probing scale $\mu^2$ dependence of $\hat q$. 
The $x_B$ and $Q^2$ dependences of $\hat q(x_B, Q^2)$ are shown in Fig. \ref{fig:qhat-xq}. The nuclear medium modification is more pronounced in small-$x$ region which is encoded in the negative power $\alpha$ in Eq.~\eqref{eq-pramt}. This is consistent with the usual small-$x$ or gluon saturation physics, which suggests that the scattering strength or gluon density increases in the small-$x$ region~\cite{Iancu:2003xm}. On the other hand, we find that the behavior of nuclear modification in the large-$x$ region, in particularly from DIS and SIDIS data, leads to a negative power $\beta$ for $(1-x)$, and indicates an enhancement of nuclear power correction in the large-$x$ region. This seems to be consistent with the theoretical consideration as advocated in~\cite{Dokshitzer:1995qm,Brodsky:2000zu}. Finally we find a weak scale $\mu^2$ dependence of $\hat q$. This is expected, as the logarithmic scale dependence is generally mild. As one can see from Fig.~\ref{fig:fit}, the theory does give a good description of all the data sets. 
\bef
\psfig{file=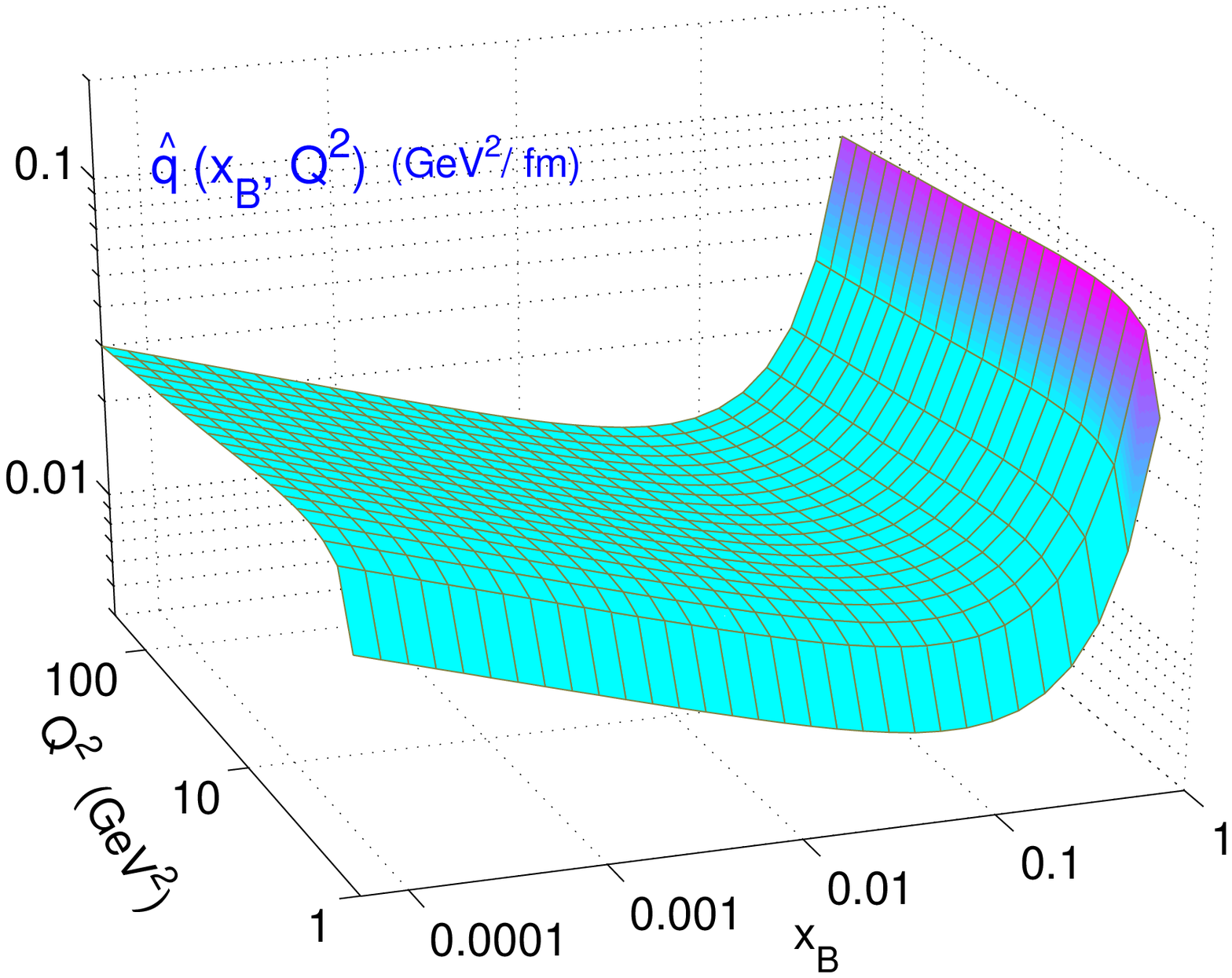, width=2.6in}
\caption{The extracted $\hat q$ as functions of Bjorken $x_B$ and scale $Q^2$.}
\label{fig:qhat-xq}
\eef
Nevertheless, we have to bear in mind that the fitted data has limited kinematic coverage, thus the extrapolation to small and large-$x$ regions needs to be further tested by future measurements in eA and pA collisions. The overall goodness $\chi^2/d.o.f = 1.21$ of our global fit provides informative value of $\hat q$ for cold nuclear matter in the intermediate $x_B$ and $Q^2$ region. 

One can further infer the jet energy dependence of $\hat q$, which is another aspect of $\hat q$ under active investigation in the community. To do that, one realizes in DIS process that $x_B = Q^2/2m\nu$ in the target rest frame, with $\nu$ the initial jet energy. Thus the $x_B$ dependence of $\hat q$ can be extended to study the jet energy dependence of $\hat q$. In the intermediate $x_B$ and $Q^2$ region, the determined parametrization gives us that $\hat q$ increases with an increasing jet energy $\nu$. Such a behavior is consistent with the expectations of jet quenching~\cite{Zhou:2019gqk,CasalderreySolana:2007sw}. We expect the determined parametrization form to have significant impact in precise extraction of fundamental property of QGP from jet quenching data, such as the jet energy dependence as discussed above.

%%%%%%%%%%%%%%%%%%%%%%%%%%%%%%%%%%%%%%%%%%%%%%%
{\it \ Summary. }\
We carried out the first global analysis of jet transport coefficient $\hat q$ for cold nuclear matter. In particular, we include the world data of transverse momentum broadening in SIDIS from HERMES, DY from SPS and FNAL, $\Upsilon$ from FNAL, and $J/\psi$ from FNAL, RHIC and LHC, comprising a total of 215 data points from 8 data sets. To check the universality of medium property, we first fit to all data sets with a constant $\hat q$ as commonly used in the literature. The failure of the fit strongly hints the kinematic and scale dependence of $\hat q$ and motivates us with a more advanced parametrization form. We then perform further global fit by assuming scale and $x$ dependence of $\hat q$ as shown in Eq. (\ref{eq-pramt}). The fitted parametrization form of $\hat q$ can describe the data very well with an overall $\chi^2/d.o.f = 1.21$. Our findings manifestly consolidate the universality of medium property and provide a data driven evidence of a scale and Bjorken-$x$ dependent $\hat q$. This parametrization form can be further extended for precise understanding of jet quenching phenomenon in relativistic heavy ion collisions. 

%%%%%%%%%%%%%%%%%%%%%%%%%%%%%%%%%%%%%%%%%%%%%%%
{\it \ Acknowledgments. }\
P.R. is supported by China Postdoctoral Science Foundation under project No. 2019M652929, Z.K. is supported by the National Science Foundation in US under Grant No. PHY-1720486, E.W., H.X. and B. Z. are supported by NSFC of China under Project No. 11435004.

%%%%%%%%%%%%%%%%%%%%%%%%%%%%%%%%%%%%%%%%%%%%%%%%%%%%%%%%%%%%

 %%%%%%%%%%%%%%
%\bibliographystyle{h-physrev5}   
%\bibliography{biblio}

\begin{thebibliography}{10}

\bibitem{Burke:2013yra}
JET, K.~M. Burke {\em et~al.},
\newblock Phys. Rev. {\bf C90}, 014909 (2014), arXiv:1312.5003.
%%CITATION = ARXIV:1312.5003;%%

\bibitem{Baier:1996sk}
R.~Baier, Y.~L. Dokshitzer, A.~H. Mueller, S.~Peigne, and D.~Schiff,
\newblock Nucl. Phys. {\bf B484}, 265 (1997), arXiv:hep-ph/9608322.
%%CITATION = HEP-PH/9608322;%%

\bibitem{Chen:2011vt}
X.-F. Chen, T.~Hirano, E.~Wang, X.-N. Wang, and H.~Zhang,
\newblock Phys. Rev. {\bf C84}, 034902 (2011), arXiv:1102.5614.
%%CITATION = ARXIV:1102.5614;%%

\bibitem{Majumder:2011uk}
A.~Majumder and C.~Shen,
\newblock Phys. Rev. Lett. {\bf 109}, 202301 (2012), arXiv:1103.0809.
%%CITATION = ARXIV:1103.0809;%%

\bibitem{Zhou:2019gqk}
F.-C. Zhou, G.-L. Ma, and Y.-G. Ma,
\newblock (2019), arXiv:1902.00729.
%%CITATION = ARXIV:1902.00729;%%

\bibitem{Xie:2019oxg}
M.~Xie, S.-Y. Wei, G.-Y. Qin, and H.-Z. Zhang,
\newblock (2019), arXiv:1901.04155.
%%CITATION = ARXIV:1901.04155;%%

\bibitem{Ma:2018swx}
G.-Y. Ma, W.~Dai, B.-W. Zhang, and E.-K. Wang,
\newblock Eur. Phys. J. {\bf C79}, 518 (2019), arXiv:1812.02033.
%%CITATION = ARXIV:1812.02033;%%

\bibitem{Chen:2016vem}
L.~Chen, G.-Y. Qin, S.-Y. Wei, B.-W. Xiao, and H.-Z. Zhang,
\newblock Phys. Lett. {\bf B773}, 672 (2017), arXiv:1607.01932.
%%CITATION = ARXIV:1607.01932;%%

\bibitem{Andres:2016iys}
C.~Andres, N.~Armesto, M.~Luzum, C.~A. Salgado, and P.~Zurita,
\newblock Eur. Phys. J. {\bf C76}, 475 (2016), arXiv:1606.04837.
%%CITATION = ARXIV:1606.04837;%%

\bibitem{Luo:1993ui}
M.~Luo, J.-w. Qiu, and G.~F. Sterman,
\newblock Phys. Rev. {\bf D49}, 4493 (1994).
%%CITATION = PHRVA,D49,4493;%%

\bibitem{Qiu:2003vd}
J.-w. Qiu and I.~Vitev,
\newblock Phys. Rev. Lett. {\bf 93}, 262301 (2004), arXiv:hep-ph/0309094.
%%CITATION = HEP-PH/0309094;%%

\bibitem{Qiu:2004da}
J.-w. Qiu and I.~Vitev,
\newblock Phys. Lett. {\bf B632}, 507 (2006), arXiv:hep-ph/0405068.
%%CITATION = HEP-PH/0405068;%%

\bibitem{Kang:2013raa}
Z.-B. Kang, E.~Wang, X.-N. Wang, and H.~Xing,
\newblock Phys. Rev. Lett. {\bf 112}, 102001 (2014), arXiv:1310.6759.
%%CITATION = ARXIV:1310.6759;%%

\bibitem{Kang:2014ela}
Z.-B. Kang, E.~Wang, X.-N. Wang, and H.~Xing,
\newblock Phys. Rev. {\bf D94}, 114024 (2016), arXiv:1409.1315.
%%CITATION = ARXIV:1409.1315;%%

\bibitem{Kang:2016ron}
Z.-B. Kang, J.-W. Qiu, X.-N. Wang, and H.~Xing,
\newblock Phys. Rev. {\bf D94}, 074038 (2016), arXiv:1605.07175.
%%CITATION = ARXIV:1605.07175;%%

\bibitem{Blaizot:2014bha}
J.-P. Blaizot and Y.~Mehtar-Tani,
\newblock Nucl. Phys. {\bf A929}, 202 (2014), arXiv:1403.2323.
%%CITATION = ARXIV:1403.2323;%%

\bibitem{Iancu:2014kga}
E.~Iancu,
\newblock JHEP {\bf 10}, 095 (2014), arXiv:1403.1996.
%%CITATION = ARXIV:1403.1996;%%

\bibitem{Liou:2013qya}
T.~Liou, A.~H. Mueller, and B.~Wu,
\newblock Nucl. Phys. {\bf A916}, 102 (2013), arXiv:1304.7677.
%%CITATION = ARXIV:1304.7677;%%

\bibitem{Qiu:2001hj}
J.-w. Qiu and G.~F. Sterman,
\newblock Int. J. Mod. Phys. {\bf E12}, 149 (2003), arXiv:hep-ph/0111002.
%%CITATION = HEP-PH/0111002;%%

\bibitem{Qiu:2005ki}
J.-W. Qiu,
\newblock Eur. Phys. J. {\bf C43}, 239 (2005), arXiv:hep-ph/0507268.
%%CITATION = HEP-PH/0507268;%%

\bibitem{Airapetian:2009jy}
HERMES, A.~Airapetian {\em et~al.},
\newblock Phys. Lett. {\bf B684}, 114 (2010), arXiv:0906.2478.
%%CITATION = ARXIV:0906.2478;%%

\bibitem{Alde:1991sw}
D.~M. Alde {\em et~al.},
\newblock Phys. Rev. Lett. {\bf 66}, 2285 (1991).
%%CITATION = PRLTA,66,2285;%%

\bibitem{Leitch:1995yc}
M.~J. Leitch {\em et~al.},
\newblock Phys. Rev. {\bf D52}, 4251 (1995).
%%CITATION = PHRVA,D52,4251;%%

\bibitem{Vasilev:1999fa}
NuSea, M.~A. Vasilev {\em et~al.},
\newblock Phys. Rev. Lett. {\bf 83}, 2304 (1999), arXiv:hep-ex/9906010.
%%CITATION = HEP-EX/9906010;%%

\bibitem{McGaughey:1999mq}
P.~L. McGaughey, J.~M. Moss, and J.~C. Peng,
\newblock Ann. Rev. Nucl. Part. Sci. {\bf 49}, 217 (1999),
  arXiv:hep-ph/9905409.
%%CITATION = HEP-PH/9905409;%%

\bibitem{Peng:1999gx}
J.-C. Peng,
\newblock AIP Conf. Proc. {\bf 494}, 503 (1999), arXiv:hep-ph/9912371.
%%CITATION = HEP-PH/9912371;%%

\bibitem{Adare:2012qf}
PHENIX, A.~Adare {\em et~al.},
\newblock Phys. Rev. {\bf C87}, 034904 (2013), arXiv:1204.0777.
%%CITATION = ARXIV:1204.0777;%%

\bibitem{Adam:2015jsa}
ALICE, J.~Adam {\em et~al.},
\newblock JHEP {\bf 11}, 127 (2015), arXiv:1506.08808.
%%CITATION = ARXIV:1506.08808;%%

\bibitem{Bordalo:1987cr}
NA10, P.~Bordalo {\em et~al.},
\newblock Phys. Lett. {\bf B193}, 373 (1987).
%%CITATION = PHLTA,B193,373;%%

\bibitem{Qiu:1990xxa}
J.-w. Qiu and G.~F. Sterman,
\newblock Nucl. Phys. {\bf B353}, 105 (1991).
%%CITATION = NUPHA,B353,105;%%

\bibitem{Qiu:1990xy}
J.-w. Qiu and G.~F. Sterman,
\newblock Nucl. Phys. {\bf B353}, 137 (1991).
%%CITATION = NUPHA,B353,137;%%

\bibitem{Luo:1994np}
M.~Luo, J.-w. Qiu, and G.~F. Sterman,
\newblock Phys. Rev. {\bf D50}, 1951 (1994).
%%CITATION = PHRVA,D50,1951;%%

\bibitem{Luo:1992fz}
M.~Luo, J.-w. Qiu, and G.~F. Sterman,
\newblock Phys. Lett. {\bf B279}, 377 (1992).
%%CITATION = PHLTA,B279,377;%%

\bibitem{Adams:1992nf}
E665, M.~R. Adams {\em et~al.},
\newblock Phys. Rev. Lett. {\bf 68}, 3266 (1992).
%%CITATION = PRLTA,68,3266;%%

\bibitem{Adams:1995is}
E665, M.~R. Adams {\em et~al.},
\newblock Z. Phys. {\bf C67}, 403 (1995), arXiv:hep-ex/9505006.
%%CITATION = HEP-EX/9505006;%%

\bibitem{Guo:1998rd}
X.-f. Guo,
\newblock Phys. Rev. {\bf D58}, 114033 (1998), arXiv:hep-ph/9804234.
%%CITATION = HEP-PH/9804234;%%

\bibitem{CasalderreySolana:2007sw}
J.~Casalderrey-Solana and X.-N. Wang,
\newblock Phys. Rev. {\bf C77}, 024902 (2008), arXiv:0705.1352.
%%CITATION = ARXIV:0705.1352;%%

\bibitem{Fries:2002mu}
R.~J. Fries,
\newblock Phys. Rev. {\bf D68}, 074013 (2003), arXiv:hep-ph/0209275.
%%CITATION = HEP-PH/0209275;%%

\bibitem{Kang:2008us}
Z.-B. Kang and J.-W. Qiu,
\newblock Phys. Rev. {\bf D77}, 114027 (2008), arXiv:0802.2904.
%%CITATION = ARXIV:0802.2904;%%

\bibitem{Kang:2012am}
Z.-B. Kang and J.-W. Qiu,
\newblock Phys. Lett. {\bf B721}, 277 (2013), arXiv:1212.6541.
%%CITATION = ARXIV:1212.6541;%%

\bibitem{GolecBiernat:1998js}
K.~J. Golec-Biernat and M.~Wusthoff,
\newblock Phys. Rev. {\bf D59}, 014017 (1998), arXiv:hep-ph/9807513.
%%CITATION = HEP-PH/9807513;%%

\bibitem{Dokshitzer:1995qm}
Y.~L. Dokshitzer, G.~Marchesini, and B.~R. Webber,
\newblock Nucl. Phys. {\bf B469}, 93 (1996), arXiv:hep-ph/9512336.
%%CITATION = HEP-PH/9512336;%%

\bibitem{Brodsky:2000zu}
S.~J. Brodsky,
\newblock {Dynamical higher twist and high x phenomena: A Window to quark quark
  correlations in QCD},
\newblock in {\em {Workshop on Nucleon Structure in High x-Bjorken Region
  (HiX2000) Philadelphia, Pennsylvania, March 30-April 1, 2000}}, 2000,
  arXiv:hep-ph/0006310.
%%CITATION = HEP-PH/0006310;%%

\bibitem{James:1975dr}
F.~James and M.~Roos,
\newblock Comput. Phys. Commun. {\bf 10}, 343 (1975).
%%CITATION = CPHCB,10,343;%%

\bibitem{Guo:2000nz}
X.-f. Guo and X.-N. Wang,
\newblock Phys. Rev. Lett. {\bf 85}, 3591 (2000), arXiv:hep-ph/0005044.
%%CITATION = HEP-PH/0005044;%%

\bibitem{Zhang:2003yn}
B.-W. Zhang and X.-N. Wang,
\newblock Nucl. Phys. {\bf A720}, 429 (2003), arXiv:hep-ph/0301195.
%%CITATION = HEP-PH/0301195;%%

\bibitem{Xing:2011fb}
H.~Xing, Y.~Guo, E.~Wang, and X.-N. Wang,
\newblock Nucl. Phys. {\bf A879}, 77 (2012), arXiv:1110.1903.
%%CITATION = ARXIV:1110.1903;%%

\bibitem{Dulat:2015mca}
S.~Dulat {\em et~al.},
\newblock Phys. Rev. {\bf D93}, 033006 (2016), arXiv:1506.07443.
%%CITATION = ARXIV:1506.07443;%%

\bibitem{Sutton:1991ay}
P.~J. Sutton, A.~D. Martin, R.~G. Roberts, and W.~J. Stirling,
\newblock Phys. Rev. {\bf D45}, 2349 (1992).
%%CITATION = PHRVA,D45,2349;%%

\bibitem{deFlorian:2007aj}
D.~de~Florian, R.~Sassot, and M.~Stratmann,
\newblock Phys. Rev. {\bf D75}, 114010 (2007), arXiv:hep-ph/0703242.
%%CITATION = HEP-PH/0703242;%%

\bibitem{Stump:2001gu}
D.~Stump {\em et~al.},
\newblock Phys. Rev. {\bf D65}, 014012 (2001), arXiv:hep-ph/0101051.
%%CITATION = HEP-PH/0101051;%%

\bibitem{Pumplin:2000vx}
J.~Pumplin, D.~R. Stump, and W.~K. Tung,
\newblock Phys. Rev. {\bf D65}, 014011 (2001), arXiv:hep-ph/0008191.
%%CITATION = HEP-PH/0008191;%%

\bibitem{Iancu:2003xm}
E.~Iancu and R.~Venugopalan,
\newblock {The Color glass condensate and high-energy scattering in QCD},
\newblock in {\em Quark-gluon plasma 4}, edited by R.~C. Hwa and X.-N. Wang,
  pp. 249--3363, 2003, arXiv:hep-ph/0303204.
%%CITATION = HEP-PH/0303204;%%

\end{thebibliography}
 
\end{document}